\documentclass{article}
\usepackage{arxiv}
\usepackage{mathtools}
\usepackage{xcolor}
\definecolor{Q}{HTML}{008000}
\definecolor{L}{HTML}{400080}
\usepackage{hyperref}

\usepackage{authblk}

\begin{document}
\newcommand{\QUAD}[1]{\textcolor{Q}{#1}}
\newcommand{\LIN}[1]{\textcolor{L}{#1}}

%\newcommand{\L}[1]{\textcolor{L}{#1}}

\title{Designing Unit Ising Models for Logic Gate Simulation through Integer Linear Programming}

\author[1]{Shunsuke Tsukiyama}
\author[1]{Koji Nakano}
\author[1]{Xiaotian Li}
\author[1]{Yasuaki Ito}
\author[2]{Takumi Kato}
\author[2]{Yuya Kawamata}
%\author{S. Tsukiyama}
%\email{webmaster@marysville-ohio.com}

\affil[1]{Graduate School of Advanced Science and Engineering, Hiroshima University, Kagamiyama 1-4-1, Higashihiroshima , Hiroshima, 739-8527, Japan}
 
\affil[2]{Research and Development Headquarters, NTT DATA Group Corporation ,Toyosu Center Bldg, Annex, 3-9, Toyosu 3-chome, Koto-ku, Tokyo, 135-8671, Japan}

\date{}
\maketitle
\begin{abstract}
An Ising model is defined by a quadratic objective function known as the Hamiltonian, composed of spin variables that can take values of either $-1$ or $+1$.
The goal is to assign spin values to these variables in a way that minimizes the value of the Hamiltonian.
Ising models are instrumental in tackling many combinatorial optimization problems, leading to significant research in developing solvers for them. 
Notably, D-Wave Systems has pioneered the creation of quantum annealers, programmable solvers based on quantum mechanics, for these models.
This paper introduces unit Ising models, where all non-zero coefficients of linear and quadratic terms are either $-1$ or $+1$.
Due to the limited resolution of quantum annealers, unit Ising models are more suitable for quantum annealers to find optimal solutions.
We propose a novel design methodology for unit Ising models to simulate logic circuits computing Boolean functions through integer linear programming.
By optimizing these Ising models with quantum annealers, we can compute Boolean functions and their inverses.
With a fixed unit Ising model for a logic circuit, we can potentially design Application-Specific Unit Quantum Annealers (ASUQAs) for computing the inverse function,
which is analogous to Application-Specific Integrated Circuits (ASICs) in digital circuitry.
For instance, if we apply this technique to a multiplication circuit, we can design an ASUQA for factorization of two numbers.
Our findings suggest a powerful new method for compromising the RSA cryptosystem by leveraging ASUQAs in factorization.
\end{abstract}


\section{Introduction}

\subsection{Quantum annealers and unit Ising models}
\begin{figure}
\centering
\includegraphics[scale=1]{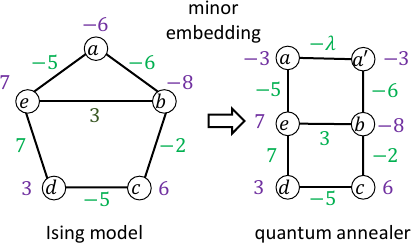}
\caption{Illustrating an Ising model with five spin variables as a weighted graph and its minor-embedding into a $3\times 2$-qubit cell quantum annealer. }
\label{fig:ising}
\end{figure}

\emph{An Ising model} is defined by a quadratic objective function known as \emph{the Hamiltonian}, which comprises spin variables that can assume values of either $-1$ or $+1$.
A simple representation of a 5-variable Hamiltonian, denoted as $H(a, b, c, d, e)$, is illustrated below:
\begin{align*}
H(a,b,c,d,e)&=\QUAD{-6ab-5ae-2bc+3be-5cd+7de}
 \LIN{-6a-8b+6c+3d+7e}
\end{align*}
For clarity, this paper differentiates between the quadratic and linear terms of Hamiltonians using distinct colors.
Within the realm of quantum computing, the coefficients of linear and quadratic terms are commonly referred to as \emph{the bias} and \emph{interaction strengths}, respectively.
The primary objective is to find a spin value assignment for the variables that minimizes the Hamiltonian's value.
For example, the aforementioned Hamiltonian $H(a, b, c, d, e)$ reaches its minimum value of $-34$ when the spin variables are assigned values of $[+1, +1, -1, -1, +1]$.
An Ising model can be visualized through a weighted graph, where each node symbolizes a variable, and the weights of nodes and edges correspond to the coefficients of linear and quadratic terms, respectively.
Figure~\ref{fig:ising} demonstrates the weighted graph representation of $H(a, b, c, d, e)$.
It is a well-established fact, as noted in~\cite{Murty87,Lucas14}, that finding an optimal spin value assignment for an Ising model is an NP-hard problem.
This implies that as long as P$\neq$NP, no algorithm, whether sequential or parallel and regardless of the polynomial hardware resources available,
can solve this problem for any Ising model in polynomial time.
Furthermore, any problem within NP can be efficiently reduced to an optimization problem of an Ising model.

\begin{figure}
\centering
\includegraphics[scale=1]{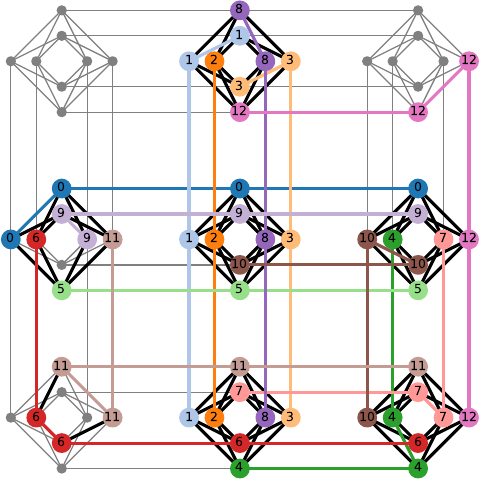}
\caption{Illustrating a minor embedding of an Ising model with $K_{13}$ topology into the Chimera graph topology with $3\times 3$ $K_{4,4}$'s}
\label{fig:chimera}
\end{figure}

Quantum annealing, which finds solutions that minimize the Hamiltonian of Ising models based on the principles of quantum mechanics, has been proposed~\cite{Kadowaki98}.
D-Wave Systems has been a pioneer in the development of quantum annealers, exemplified by their innovative D-Wave 2000Q model~\cite{McGeoch19}.
This quantum annealer features 2048 qubit cells, interconnected according to the Chimera graph topology.
This topology consists of $16\times 16$ blocks of $K_{4,4}$ graphs, each representing a complete bipartite graph with nodes $(4,4)$. 
Figure~\ref{fig:chimera} illustrates the Chimera graph, here depicted with $3\times 3$ blocks of $K_{4,4}$ graphs.
Following this, D-Wave Systems launched an even more sophisticated quantum annealer, the D-Wave Advantage~\cite{Advantage}.
This model comprises 5760 qubit cells, interconnected using the Pegasus graph topology, a complex network of 5760 nodes~\cite{DWaveAdvantage19}.
Given that the topology of the Ising models solvable by a quantum annealer is predetermined, any Ising model intended for this solver must first undergo a transformation known as \emph{minor embedding} to align with the annealer's topology.
Figure~\ref{fig:ising} demonstrates minor embedding through a simple example, where a function $H(a,b,c,d,e)$ is embedded into a grid of 6 qubit cells arranged in a $3\times 2$ formation.
The Ising model embedded in these 6 qubit cells is defined as follows:
\begin{align*}
H(a,a',b,c,d,e)&= \QUAD{-6a'b-5ae-2bc+3be-5cd+7de-\lambda aa'}\\
&\quad \LIN{-3a-3a'-8b+6c+3d+7e}
\end{align*}
Here, node $a$ from $H(a,b,c,d,e)$ is mapped to qubit cells $a$ and $a'$, with the interconnecting edge assigned a weight of $-\lambda$ ($\lambda>0$).
During the quantum annealing process, the qubit cells adopt spin states that minimize the value of $H(a,a',b,c,d,e)$.
Significantly, $-\lambda aa' = -\lambda$ when $a=a'$, and $+\lambda$ when $a\neq a'$.
This parameter $\lambda$, known as \emph{chain strength} in D-Wave quantum annealers, must be chosen with a sufficiently high value to ensure that the quantum annealing process identifies solutions where $a=a'$ as viable.
Since $H(a,b,c,d,e)=H(a,a',b,c,d,e)+\lambda$ is true for any feasible solution set $a, a', b, c, d, e$, we can determine the solution for the original Ising model $H(a,b,c,d,e)$ through quantum annealing of $H(a,a',b,c,d,e)$, provided that $\lambda$ is adequately large.

For Ising models with dense graph topologies, minor embedding might require mapping each node to multiple qubit cells\cite{Nakano23-dual}.
Figure~\ref{fig:chimera} demonstrates this with a $K_{13}$ (13-node complete graph) embedded into a Chimera graph of $3\times 3$ $K_{4,4}$ blocks.
This embedding, achievable through D-Wave Systems' minorminer tool~\cite{Cai14}, involves mapping each Ising model node to 3 to 5 qubit cells in the Chimera graph, connected by chains of edges.

When using quantum annealers, such as the D-Wave Advantage, it is crucial to address the limitations in coefficient resolution.
The range of coefficients in programmed Ising models is confined to $[-4.0, +4.0]$ for linear terms and $[-1.0, +1.0]$ for quadratic terms.
Ising models featuring coefficients beyond these ranges require scaling to adhere to these constraints.
However, since quantum annealing operates on an analog basis, small or closely-valued coefficients may be obscured by flux noise, as detailed in~\cite{Zaborniak21}.
Practically, the resolution of the D-Wave Advantage approximates to 5 bits, with values below $2^{-5} \approx 0.03$ potentially being overwhelmed by noise.
To maintain clear resolution, it is assumed that all coefficients in Ising models are integers, with the smallest non-zero coefficient set to 1.
This assumption results in a pitch width of $1 - (-1) = 2$ in outcome values, considering that spin variables and their products can only be $-1$ or $+1$.
The necessary resolution for scaling Ising models for quantum annealers is determined by the maximum absolute coefficient value.
For models with large maximum absolute values, it is essential to divide coefficients by a substantial constant, which can challenge the annealer’s ability to identify optimal solutions.
To circumvent these limitations, we suggest adopting \emph{unit Ising models}, where all non-zero coefficients are designated as \emph{unit values}, namely $-1$ or $+1$.
This approach eliminates the need for scaling, making the quantum annealing process more robust against noise.

\subsection{Our contribution}
\begin{figure}
\centering
\includegraphics[scale=1]{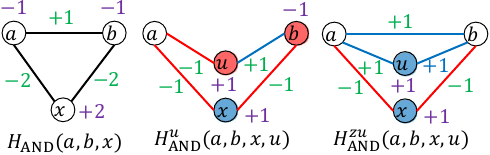}
\caption{An Ising model $H_{\rm AND}(a,b,x)$, a unit Ising model $H^{u}_{\rm AND}(a,b,x,u)$ and a zero-input unit Ising model $H^{zu}_{\rm AND}(a,b,x,u)$ for simulating the AND gate}
\label{fig:and}
\end{figure}

The primary contribution of this paper is to introduce \emph{unit Ising models} and present a methodology of designing them
for simulating logic circuits.
Throughout this paper, we use the convention where the false (or 0) and true (or 1) values in logic circuits and Boolean functions are represented by spin values $-1$ and $+1$ in Ising models, respectively.
To facilitate readers' understanding, we provide a clear example of simulating a logic circuit using an Ising model for a single AND gate. 
The Ising model, denoted as $H_{\rm AND}(a, b, x)$, simulates an AND gate with inputs $a$ and $b$, and output $x$:
\begin{align*}
H_{\rm AND}(a, b, x) &=\QUAD{ab - 2ax - 2bx}~\LIN{- a - b + 2x}
\end{align*}
The Ising model achieves its minimum value of $-3$ when $x = a \wedge b$ is satisfied, and $H(a, b, x) \geq -2$ in other cases.
Thus, when the spin values of $a$ and $b$ are fixed, the value of $x$ will be $a \wedge b$ in the optimal solution obtained
by \emph{forward simulation} using a quantum annealer.
Conversely, when the spin value of $x$ is fixed, the values of $a$ and $b$ will align with $x = a \wedge b$ through \emph{backward simulation}.
For example, if $x=-1$ (false), then $(a,b)$ will be one of $(-1,-1)$, $(-1,+1)$, or $(+1,-1)$.
Note that $H_{\rm AND}(a, b, x)$ is not a unit Ising model and it is not feasible to design a unit Ising model with three variables for the AND gate.
However, unit Ising models for simulating an AND gate can be designed using an ancillary variable $u$, as follows:
\begin{align*}
H^{u}_{\rm AND}(a, b, x, u) &=\QUAD{-au - ax + bu - bx}~\LIN{- b - u + x}\\
H^{zu}_{\rm AND}(a, b, x, u) &= \QUAD{ab + au - ax + bu - bx}~\LIN{+ u + x}
\end{align*}
The unit Ising model $H^{u}{\rm AND}(a, b, x, u)$ simulates an AND gate, as either $H^{u}{\rm AND}(a, b, x, -1)$ or $H^{u}{\rm AND}(a, b, x, +1)$ reaches the minimum value of $-3$
when $x=a \wedge b$, and $H^{u}{\rm AND}(a, b, x, u) \geq -2$ when $x\neq a \wedge b$.
The unit Ising model $H^{zu}{\rm AND}(a, b, x, u)$ also simulates an AND gate, as it possesses the same characteristics.
This model is called \emph{zero-input}, because it lacks linear terms for $a$ and $b$, which correspond to the inputs of an AND gate.
As we will explain later, we introduce zero-input property to enable the removal of edges between input and output nodes, reducing the size of Ising models for simulating logic circuits.
Figure~\ref{fig:and} illustrates the corresponding weighted graphs for $H(a,b,x)$, $H^{u}{\rm AND}(a, b, x, u)$, and $H^{zu}{\rm AND}(a, b, x, u)$.
In these unit Ising models, coefficients $-1$ and $+1$ are depicted in red and blue, respectively,
while a coefficient 0 of linear terms is uncolored.
Although $H^{zu}{\rm AND}(a, b, x, u)$ is a unit Ising model, it includes more quadratic terms (or edges) than $H^{u}_{\rm AND}(a, b, x, u)$.
We can select the most suitable Ising model from these two when designing a logic circuit.

We demonstrate that  zero-input unit Ising models, which simulate logic gates, can be developed through integer linear programming.
Our method involves formulating integer linear programming models to derive Ising models and obtaining solutions with solvers like Gurobi Optimizer~\cite{Gurobi}.
These solutions enable the construction of zero-input unit Ising models for simulating logic gates.
Furthermore, our approach allows for the verification of the existence of Ising models.
For instance, we show that a unit Ising model for an AND gate, without ancillary variables, does not exist.

\begin{figure}
\centering
\includegraphics[scale=1]{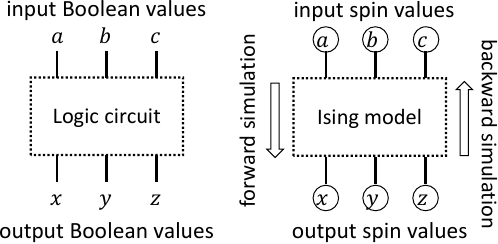}
\caption{A logic circuit with inputs $a, b, c$ and outputs $x,y,z$ and the unit Ising model simulating it.} %この図のCaptionはほかのCaptionと違ってピリオドがあります
\label{fig:circuit}
\end{figure}

Combinational logic circuits consist of multiple logic gates.
Therefore, by combining Ising models that simulate individual logic gates, we can design an Ising model to emulate a combinational logic circuit.
In an Ising model, nodes (or variables) are assigned spin values corresponding to their inputs.
Specifically, an Ising model designed to simulate a combinational logic circuit includes variables representing both the circuit's inputs and outputs.
For example, Figure~\ref{fig:circuit} illustrates a logic circuit with inputs $a, b, c$ and outputs $x, y, z$, along with its simulating Ising model.
By assigning Boolean values to inputs $a, b, c$, the logic circuit determines the Boolean outcomes for outputs $x, y, z$.
To mimic this computation in the Ising model, the spin values of variables $a, b, c$ are fixed, and the optimal solutions for variables $x, y, z$ are obtained through quantum annealing.
This process, termed \emph{forward simulation}, computes the output Boolean values from given inputs.

It makes no sense to use quantum annealers for the forward simulation of logic circuits, as current digital devices can efficiently perform this task using logic circuits.
However, \emph{backward simulation} using an Ising model introduces a compelling application.
Through this process, we can determine the inputs for a logic circuit by finding the optimal solution of the corresponding Ising model.
Essentially, for a Boolean function $f$ computed by a logic circuit, backward simulation enables us to compute its inverse $f^{-1}$ using the Ising model.
In computer science, a function $f$ is termed \emph{one-way} if $f$ can be easily computed, whereas its inverse $f^{-1}$ is significantly more challenging to compute~\cite{Levin03}.
While the existence of one-way functions remains a conjecture, certain functions are considered as potential one-way functions.
For example, the multiplication function $f$ is straightforward, but its inverse, the factoring function $f^{-1}$, is notoriously difficult, a principle underlying the RSA cryptosystem~\cite{RSA}.
This methodology in designing Ising models for both forward and backward simulations of logic circuits implies that for any one-way function $f$ implementable in a logic circuit,
its inverse $f^{-1}$ can be computed by finding optimal solutions for the Ising models.
Using this methodology, we present a zero-input unit Ising model for multiplication and factorization.

We also discuss \emph{the area complexity of the Ising model}. To find the optimal solution for an Ising model using quantum annealers, qubit cells and interconnects are arranged in a 2-dimensional plane. The hardware complexity of a quantum annealer can be evaluated by the size of the area necessary to implement this Ising model. It is sensible to adopt the assumptions used in the VLSI complexity model \cite{Ullman84}. More specifically, we assume that qubit cells occupy $O(1)$ area, and the width of interconnects is $O(1)$. Under these assumptions, we will show that our Ising model for multiplication and factorization of $n$-bit numbers can be laid out in a 2-dimensional area of size $O(n^2)$ 
Since any Ising model for the multiplication and factorization for $n$-bit numbers requires $\Omega(n^2)$ area, it is \emph{area-optimal}.
To our knowledge, this paper is the first to introduce the concept of area complexity in Ising models and demonstrate their optimality.

This approach illustrates the potential for using \emph{an Application-Specific Unit Quantum Annealer (ASUQA)}, which is tailored for a particular one-way function $f$ to facilitate both its forward and backward simulations.
For instance, an ASUQA designed for multiplication and factorization features a predetermined Ising model topology, with internal qubit cells and interconnects assigned fixed biases of either $-1$, $0$, or $+1$ and interaction strengths of either $-1$ or $+1$.
Through backward simulation, such an ASUQA enables the factorization into two prime numbers via quantum annealing.
Alternatively, \emph{General-Purpose Quantum Annealers (GPQAs)}, like the D-Wave Advantage, can be employed for the same objective.
However, this method requires minor-embedding to configure the Ising model, necessitating a substantial strength parameter $\lambda$.
Minor-embedding significantly increases the number of qubit cells and interconnections, along with the required chain strength, which complicates the optimization process for GPQAs.
Conversely, an ASUQA, designed without the need for minor-embedding, is inherently more resistant to flux noise and errors due to its unit-value coefficients.

Furthermore, the relationship between ASUQAs and GPQAs in quantum computing mirrors the distinction between ASICs (Application-Specific Integrated Circuits) and FPGAs (Field Programmable Gate Arrays) in the domain of digital electronics.
Like ASICs, ASUQAs are tailored for specific computations, making full use of the hardware resources to achieve higher efficiency for designated tasks compared to their general-purpose counterparts, GPQAs, and FPGAs.
Alternatively, one might consider utilizing a non-unit Application-Specific Quantum Annealer (ASQA), which operates on a non-unit Ising model with various biases and interaction strengths.
Nonetheless, the architecture of ASUQAs, which limits values to either $-1$ or $+1$, offers more reliability and lower costs compared to the non-unit nature of ASQAs.
This preference for digital solutions parallels the trend in classical (non-quantum) electronics, where digital VLSI technologies have become more prevalent than analog VLSI technologies~\cite{Mead89} due to their enhanced reliability and reduced production costs.

This paper is organized as follows:
Section~\ref{sec:gate} details the design of zero-input unit Ising models for simulating basic logic gates, including AND, OR, and XOR gates, using linear programming.
Section~\ref{sec:circuit} demonstrates the design of Ising models for simulating logic circuits utilizing these zero-input unit Ising models.
In Section~\ref{sec:factorization}, we present a case study on computing the inverse of one-way functions by designing an Ising model for factorization.
Section~\ref{sec:related} discusses works related to this paper.
Finally, Section~\ref{sec:concl} provides the conclusion of our work.

\section{Finding Ising models for logic gates using linear programming}
\label{sec:gate}
This section demonstrates that Ising models simulating logic gates can be obtained through integer linear programming.
We first illustrate how to derive Ising models simulating an AND gate.

Recall that spin variables assume values of $-1$ or $+1$, corresponding to the Boolean values false (or 0) and true (or 1), respectively, in logic circuits.
An Ising model incorporating three spin variables, $a$, $b$, and $x$, can be expressed as:
\begin{align*}
H(a,b,x) &=\QUAD{AB\cdot ab+AX\cdot ax+BX\cdot bx}~\LIN{+A\cdot a+B\cdot b+X\cdot x},
\end{align*}
where $AB$, $AX$, $\ldots$, $X$ represent the integer coefficients for $ab$, $ax$, $\ldots$, $x$, respectively.
The objective is to identify coefficients such that the Ising model achieves its minimum value if and only if $x = a \wedge b$.
To this end, we introduce $2^3 = 8$ constraints for the spin value assignments to $a$, $b$, $x$, ensuring that $H(a, b, x)$ reaches a minimum value $\mu$, i.e., $H(a, b, x) = \mu$ if $x = a \wedge b$ (consistent) and $H(a, b, x) \geq \mu + 1$ when $x \neq a \wedge b$ (inconsistent).
\begin{align*}
H(-1, -1, -1) &=\QUAD{+{\it AB}+{\it AX}+{\it BX}}~\LIN{-A-B-X}= \mu \\
H(-1, -1, +1) &=\QUAD{+{\it AB}-{\it AX}-{\it BX}}~\LIN{-A-B+X}\geq \mu+1\\
H(-1, +1, -1) &=\QUAD{-{\it AB}+{\it AX}-{\it BX}}~\LIN{-A+B-X}= \mu\\
H(-1, +1, +1) &=\QUAD{-{\it AB}-{\it AX}+{\it BX}}~\LIN{-A+B+X}\geq \mu+1\\
H(+1, -1, -1) &=\QUAD{-{\it AB}-{\it AX}+{\it BX}}~\LIN{+A-B-X}= \mu\\
H(+1, -1, +1) &=\QUAD{-{\it AB}+{\it AX}-{\it BX}}~\LIN{+A-B+X}\geq \mu+1\\
H(+1, +1, -1) &=\QUAD{+{\it AB}-{\it AX}-{\it BX}}~\LIN{+A+B-X}\geq \mu+1\\
H(+1, +1, +1) &=\QUAD{+{\it AB}+{\it AX}+{\it BX}}~\LIN{+A+B+X}= \mu
\end{align*}
Furthermore, we adopt objectives aimed at minimizing to derive a zero-input unit Ising model as follows:
\begin{align*}
&\textrm{MAX\_ABS:} &&\quad \max(|AB|,|AX|,|BX|,|A|,|B|,|X|),\\
&\textrm{INPUT\_NUM:} &&\quad (A\neq 0)+(B\neq 0),\\
&\textrm{QUAD\_NUM:} &&\quad (AB\neq 0)+(AX\neq 0)+(BX\neq 0).
\end{align*}
In these constraints, $(A \neq 0)$ evaluates to 1 if $A \neq 0$, and 0 otherwise.
The priority objective, MAX\_ABS, is the maximum absolute value among all coefficients, aiming for a unit Ising model. 
The second priority, INPUT\_NUM, counts the number of non-zero coefficient linear terms corresponding to inputs, seeking a zero-input Ising model.
QUAD\_NUM counts the non-zero coefficient quadratic terms.
Utilizing a integer programming solver, such as Gurobi Optimizer~\cite{Gurobi}, we find an optimal solution: $AB = 1$, $AX = BX = -2$, $A = B = -1$, $X = 2$, and $\mu = -3$, with objective values MAX\_ABS $= 2$, INPUT\_NUM $= 2$, and QUAD\_NUM $= 3$, thereby constructing the Ising model $H_{\text{AND}}(a, b, x)$ for the AND gate.
Since no solution exists with MAX\_ABS $=1$, unit Ising models simulating the AND gate with three variables are not feasible.

Given the non-unit nature of the optimal solution for the AND gate, we introduce an ancillary variable $u$ to derive a unit Ising model, $H(a, b, x, u)$, as follows:
\begin{align*}
H(a,b,x,u) &=\QUAD{AB\cdot ab+AX\cdot ax+AU\cdot au+BX\cdot bx +BU\cdot bu}\\
&\quad \QUAD{+XU\cdot xu}~\LIN{+A\cdot a+B\cdot b+X\cdot x+U\cdot u}.
\end{align*}
We use constraints such that $H(a, b, x, -1) \geq \mu$ and $H(a, b, x, +1) \geq \mu$,
and at least one of the equalities holds true if $x = a \wedge b$.
Additionally, $H(a, b, x, -1) \geq \mu + 1$ and $H(a, b, x, +1) \geq \mu + 1$ if $x \neq a \wedge b$.
For example, either $H(-1, -1, -1, -1) = \mu$ or $H(-1, -1, -1, +1) = \mu$ holds from $(-1) = (-1) \wedge (-1) $, and thus, we use the following constraint:
\begin{align*}
\QUAD{+AB+AX+AU+BX+BU+XU}~\LIN{-A-B-X-U}&=\mu && \mbox{or}\\
\QUAD{+AB+AX-AU+BX-BU-XU}~\LIN{-A-B-X+U}&= \mu
\end{align*}
Since $H(a, b, x, u)$ should not be below $\mu$, we need the following constraints:
\begin{align*}
\QUAD{+AB+AX+AU+BX+BU+XU}~\LIN{-A-B-X-U}&\geq \mu\\
\QUAD{+AB+AX-AU+BX-BU-XU}~\LIN{-A-B-X+U}&\geq \mu
\end{align*}
From the scenario where $(+1) \neq (-1) \wedge (-1)$, we employ:
\begin{align*}
\QUAD{+AB-AX+AU-BX+BU-XU}~\LIN{-A-B+X-U} & \geq \mu+1 \\
\QUAD{+AB-AX-AU-BX-BU+XU}~\LIN{-A-B+X+U} & \geq \mu+1
\end{align*}
Objectives MAX\_ABS, INPUT\_NUM, and QUAD\_NUM are applied identically as before.
By solving this integer linear programming problem, we obtain an optimal solution: $A=B=UX=0$, $AB = AU = BU =U=X= +1$, $AX = BX =-1$,and $\mu = -3$ with objective values MAX\_ABS $= 1$, INPUT\_NUM$= 0$ and QUAD\_NUM$= 5$, providing the zero-input unit Ising model $H^{zu}_{\text{AND}}(a, b, x, u)$ for the AND gate.
The computed values of $H^{zu}{\text{AND}}(a, b, x, u)$ across all potential input combinations of $a$, $b$, $x$, and $u$ are detailed in Table~\ref{tab:and}.
It is observed that $H^{zu}{\text{AND}}(a, b, x, u) \geq -2$ whenever $x \neq a \wedge b$, and either $H^{zu}{\text{AND}}(a, b, x, -1)$ or $H^{zu}{\text{AND}}(a, b, x, +1)$ achieves the minimum value of $-3$ if $x = a \wedge b$.
Consequently, all optimal solutions of $H^{zu}_{\rm AND}(a,b,x,u)$ satisfy $x= a\wedge b$ and all the other solutions do not. 
{
\setlength{\tabcolsep}{4pt}
\begin{table}
\centering
\caption{The values of $H^{u}_{\rm AND}(a,b,x)$ with $H^{zu}_{\rm AND}(a,b,x)$ for inputs $a,b,x$}
\label{tab:and}
\begin{tabular}{c|c|ccc|ccc}
$a,b,x$ & $x\stackrel{?}{=}a \wedge b$& $u$ & $H^{u}_{\rm AND}$ & $H^{zu}_{\rm AND}$ & $u$ & $H^{u}_{\rm AND}$& $H^{zu}_{\rm AND}$  \\
\hline
$-1, -1, -1$&$=$ & $-1$ & $-1$ & $-1$ &
$+1$ &\boldmath$-3$ &\boldmath$-3$ \\
$-1, -1, +1$&$\neq$&$-1$ & $+5$&$+5$ &
$+1$ & $+3$ &$+3$\\
$-1, +1, -1$&$=$&$-1$ &  \boldmath$-3$& \boldmath$-3$&
$+1$ & $-1$ & $-1$ \\
$-1, +1, +1$&$\neq$&$-1$ & $-1$&$-1$& 
$+1$ & $+1$& $+1$\\
$+1, -1, -1$&$=$&$-1$ & $+3$ & \boldmath$-3$&
$+1$ & \boldmath$-3$ & $-1$\\
$+1, -1, +1$&$\neq$&$-1$ &$+5$& $-1$&
$+1$ & $-1$ & $+1$\\
$+1,+ 1, -1$&$\neq$ &$-1$ &$+1$ & $-1$&
$+1$ &$-1$ & $+5$\\
$+1,+1, +1$&$=$&$-1$ & $-1$ & \boldmath$-3$&
$+1$ &  \boldmath$-3$ & $+3$
\end{tabular}
\end{table}
}

Ising models corresponding to the OR and XOR gates are also derived using a methodology analogous to that employed for the AND gate.
The essence of this approach lies in utilizing integer linear programming to construct models that simulate the desired logical operations.
The outcomes of this process are concisely encapsulated in Table~\ref{tab:basic}, which presents a summary of the Ising models formulated for each gate type through the application of integer linear programming techniques.
\begin{table*}[!ht]
\centering
\caption{Ising models for simulating basic logic gates}
\label{tab:basic}
\begin{tabular}{c|llc}
logic gate &\multicolumn{2}{c}{Ising model}& $\mu$\\
\hline
$x=a\wedge b$ &$H_{\rm AND}(a,b,x)$& $\QUAD{ab - 2ax - 2bx}$~$\LIN{-a - b + 2x}$ &$-3$ \\
$x=a\wedge b$ & $H^{u}_{\rm AND}(a, b, x, u$) &$\QUAD{-au - ax + bu - bx}$~$\LIN{-b - u + x}$&$-3$ \\
$x=a\wedge b$ & $H^{zu}_{\rm AND}(a,b,x,u)$ & $\QUAD{ab + au - ax + bu - bx}$~$\LIN{+u + x}$&$-3$ \\
$x=a\vee b$ &$H_{\rm OR}(a,b,x)$& $\QUAD{ab - 2ax - 2bx}$~$\LIN{+a + b - 2x}$ &$-3$\\
$x=a\vee b$ &$H^{u}_{\rm OR}(a,b,x,u)$&$\QUAD{-au - ax + bu - bx}$~$\LIN{+b + u - x}$ &$-3$\\
$x=a\vee b$ &$H^{zu}_{\rm OR}(a,b,x,u)$&$\QUAD{ab + au - ax + bu - bx}$~$\LIN{-u - x}$ &$-3$\\
$x=a\oplus b$ &$H_{\rm XOR}(a,b,x)$ & none (no feasible solution) \\
$x=a\oplus b$ &$H_{\rm XOR}(a,b,x,u)$ & $\QUAD{-ab + 2au + ax - 2bu - bx + 2ux}$~$\LIN{-a + b - 2u - x}$ &$-4$\\
$x=a\oplus b$ & $H^{zu}_{\rm XOR}(a,b,x,u,v)$& $\QUAD{-ab - au - av + bu + bv + ux - vx}$~$\LIN{-u + v - x}$&$-4$\\
\hline
$s=x\oplus y$ &$H_{\rm HA}(x,y,s,c)$ & $\QUAD{2cs - 2cx - 2cy - sx - sy + xy +2c}$~$\LIN{ + s - x - y}$&$-4$\\
$c=x\wedge y$ & $H^{zu}_{\rm HA}(x,y,s,c,u)$ & $\QUAD{cs - cx - cy + su + ux + uy + xy}$~$\LIN{+c + s + u}$ &$-4$\\
\hline
$s=x\oplus y\oplus z$& $H_{\rm FA}(x,y,z,s,c)$ & $\QUAD{2cs - 2cx - 2cy - 2cz - sx - sy - sz + xy + xz + yz}$&$-4$\\
$c={\rm maj}(x,y,z)$ & $H^{zu}_{\rm FA}(x,y,z,s,c,u)$ & $\QUAD{cs - cx - cy - cz - su - sy - ux + uy - uz + xz}$&$-4$\\
\end{tabular}
\end{table*}

\begin{figure}[!ht]
\centering
\includegraphics[scale=1]{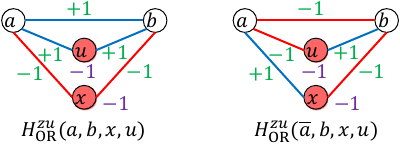}
\caption{Illustrating zero-input unit Ising models $H^{zu}_{OR}(a,b,x,u)$ and $H^{zu}_{OR}(\overline{a},b,x,u)$}
\label{fig:or}
\end{figure}

Ising models, which simulate logic gates such as AND,  
OR, and XOR, can be extended to include cases where some of the inputs and/or outputs are inverted.
This flexibility allows for the simulation of a wider array of logical operations, further demonstrating the versatility of Ising models in computational representations.
As an example, consider the modification required to simulate an OR gate with an inverted input, $x = \overline{a} \vee b$. This can be achieved by inverting the input $a$ in the Ising model, leading to:
\begin{align*}
H_{OR}(\overline{a},b,x,u) &= \QUAD{\overline{a}b + \overline{a}u - \overline{a}x+ bu - bx}\LIN{ -u - x}\\
   &= \QUAD{-ab - au + ax + bu - bx}\LIN{-u - x}.
\end{align*}
Similarly, for simulating an XNOR gate, which satisfies $x = \overline{a \oplus b}$ (equivalently, $\overline{x} = a \oplus b$), the inversion is applied to the output $x$ in the Ising model of XOR gate:
\begin{align*}
\MoveEqLeft{H_{XNOR}(a,b,x,u,v)=H_{XOR}(a,b,\overline{x},u,v)}\\
 &=\QUAD{-ab - au - av + bu + bv + u\overline{x} - v\overline{x}}\LIN{-u + v - \overline{x}}\\
 &=\QUAD{-ab - au - av + bu + bv - ux + vx}\LIN{-u + v + x}
\end{align*}
Figure~\ref{fig:or} illustrates zero-input unit Ising models $H^{zu}_{OR}(a,b,x,u)$ and $H^{zu}_{OR}(\overline{a},b,x,u)$. 
These modifications demonstrate the versatility of Ising models in representing a wide range of logical operations through inversion.
In fact, all 16 two-input logic functions~\cite{Su16}, including AND, OR, XOR, and XNOR, can be modeled by appropriately negating inputs and/or outputs.

It is worth noting that unit Ising models with fewer quadratic terms may be found if zero-input is not required.
To achieve such Ising models, we prioritize the objectives MAX\_ABS, QUAD\_NUM, and INPUT\_NUM in this order.
Through integer linear programming with these objective functions, we have obtained Ising models $H^{u}{\text{AND}}(a,b,x,u)$ and $H^{u}{\text{OR}}(a,b,x,u)$, which are not zero-input but have only 4 quadratic terms.
Table~\ref{tab:and} also presents the values of $H^{u}{\text{AND}}(a,b,x,u)$ for all possible combinations of $a$, $b$, $x$, and $u$.
However, the Ising model for XOR gate simulation obtained through this technique remains the same as $H^{zu}{\text{XOR}}(a,b,x,u,v)$, because INPUT\_NUM and QUAD\_NUM can be minimized simultaneously.

\section{Designing Ising models to simulate combinational logic circuits using unit Ising models}
\label{sec:circuit}
This section demonstrates the derivation of an Ising model for simulating combinational logic circuits.
It is clear that combinational logic circuits can be implemented using AND, OR, and XOR gates, with or without inverters.
Accordingly, we will detail the steps involved in transforming a combinational logic circuit into an equivalent Ising model.
For clarity, we design a weighted graph representing a unit Ising model.
Recall that a weighted graph has nodes and edges corresponding to spin variables and quadratic terms of a unit Ising model, respectively.
The weights of nodes and edges represent the coefficients of linear and quadratic terms.
Red and blue colors in illustrations of weighted graphs denote coefficients of $-1$ and $+1$, respectively.

\begin{figure*}[!ht]
\centering
\includegraphics[scale=1]{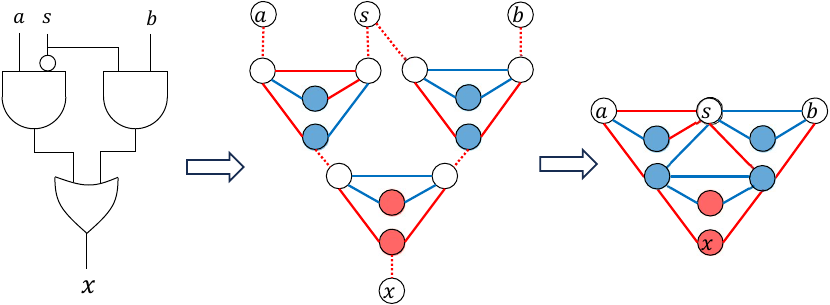}
\caption{An example of generating a zero-input unit Ising model simulating a logic circuit and simplification of it}
\label{fig:sel}
\end{figure*}

A desired unit Ising model for simulating a combinational logic circuit can be constructed as follows:
for each input and output of the circuit, the weighted graph includes a node with no color to simulate it.
Additionally, for each logic gate, we use a weighted graph of the Ising model to represent it.
Finally, for each wire in the logic circuits, the corresponding nodes in the undirected graphs are connected by a red edge representing a coefficient of $-1$.
This red edge ensures that two nodes connected by it will assume the same spin value.
Figure~\ref{fig:sel} illustrates a logic circuit designed to compute $x = (a \land \overline{s}) \lor (b \land s)$, along with a weighted graph representing the corresponding Ising model.
The weighted graph consists of four nodes: $a$, $b$, $s$, and $x$, each with no color, corresponding to the inputs and output of the logic circuit.
Additionally, there are three sets of Ising models, each with four nodes, simulating the respective gates.
In the figure, red dotted lines represent red edges simulating wires of the logic circuit.

The Ising model obtained can be utilized in two distinct ways: \emph{forward simulation} and \emph{backward simulation}.
In \emph{the forward simulation}, input values are provided to the logic circuit, and the corresponding output values are computed using the Ising model.
This allows for the analysis of the circuit's behavior when subjected to specific input values.
Conversely, in the \emph{the backward simulation}, output values of the logic circuit are given,
and the Ising model is employed to compute input values that are consistent with the provided output values.
This type of simulation aids in understanding the possible sets of inputs that could produce a particular set of outputs in the logic circuit.
We use the weighted graph in Figure~\ref{fig:sel} for explaining these simulations.
Suppose that $(a,s,b)=(+1,-1,+1)$ are assigned to three nodes as the fixed spin values for the forward simulation.
It is evident that an optimal solution of the Ising problem for the weighted graph must be consistent with the logic circuit.
If a solution (or spin value assignment to every node) is not consistent with the logic circuit, it cannot be an optimal solution.
Hence, the optimal solution of the Ising model yields $x=+1$.
Suppose that $x=-1$ is assigned for the backward simulation.
Since the optimal solution of the Ising model must be consistent with the logic circuit,
an optimal solution corresponds to one of all possible input spin value assignment
$(a,s,b)$$=$$(-1,-1,-1),$$ (-1,-1,+1),$$ (-1,+1,-1),$$(+1,+1,-1)$.

\begin{figure}[!ht]
\centering
\includegraphics[scale=1]{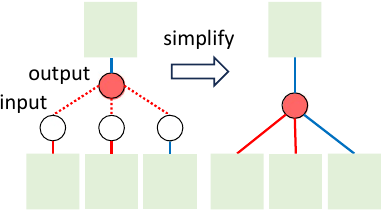}
\caption{An example of merging an output node and three input nodes of zero-input Ising models into one node}
\label{fig:simplify}
\end{figure}

It is worth mentioning that zero-input unit Ising models can always be simplified by merging two nodes connected by a red dotted edge.
Figure~\ref{fig:sel} also illustrates the simplified unit Ising model.
Since each blue output node is connected to an input node with no color, they are merged into one blue node. Such simplification is always possible for all logic circuits because:
\begin{itemize}
\item one output node is connected to one or more input nodes by red dotted edges,
\item no two output nodes are never connected by red dotted edges, and
\item all input nodes have no color, while output nodes may have color.
\end{itemize}
Thus, a connected component of red dotted edges has at most one colored node, and all nodes in the component can always be merged into one node while preserving the color. Therefore, all red dotted edges in a weighted graph can be removed by merging the two nodes. Figure~\ref{fig:simplify} illustrates an example of the merging operation for zero-input Ising models.

\section{Ising models for factorization}
\label{sec:factorization}
This section delves into an Ising model designed for the multiplication of two numbers.
The forward simulation within this Ising model facilitates the multiplication of two numbers.
In cases where the product of two prime numbers is provided, the backward simulation can identify these prime numbers as an optimal solution.

\begin{figure}[!ht]
\centering
\includegraphics[scale=1]{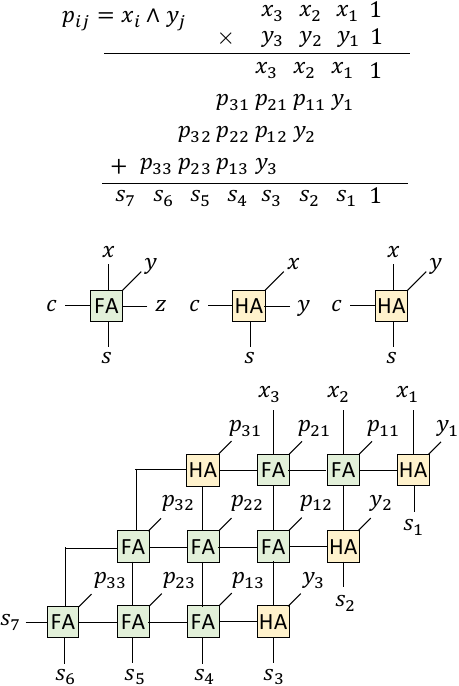}
\caption{Illustrating the binary multiplication of two 4-bit numbers $x_3x_2x_1x_0$ and $y_3y_2y_1y_0$, and the multiplication circuit simulating the binary multiplication}
\label{fig:grade-school}
\end{figure}

We review a widely-used binary multiplication circuit that computes the product of two numbers using AND gates, half adders (HAs), and full adders (FAs).
Let $X=x_{n-1}x_{n-2}\cdots x_0$ (multiplicand) and $Y=y_{n-1}y_{n-2}\cdots y_0$ (multiplier) be two $n$-bit binary numbers, and let $S=s_{2n-1}s_{2n-2}\cdots s_{0}$ be their product to be computed.
Since we are interested in factorization, we can assume $X$ and $Y$ are odd numbers, meaning $x_0=y_0=s_0=1$ holds to simplify the logic circuit and the Ising model for simulating it.
We employ a multiplication circuit based on a method known as grade-school multiplication for decimal numbers, which computes the product of the multiplicand and each digit of the multiplier, and then sums up all the properly shifted results.
Figure~\ref{fig:grade-school} illustrates the grade-school multiplication for two 4-bit numbers.
It can be simulated by a multiplication circuit using AND gates, half adders, and full adders. 
A half adder has two inputs, $x$ and $y$, and computes the sum output $s$ and the carry output $c$, satisfying $s=x\oplus y$ and $c=x\wedge y$.
A full adder has three inputs, $x$, $y$, and $z$, and computes the sum output $s$ and the carry output $c$, satisfying $s=x\oplus y\oplus z$ and $c={\rm maj}(x,y,z)$, where ${\rm maj}()$ denotes the majority function that evaluates to true when more than half of the arguments are true, and false otherwise.
The multiplication circuit has two components:
\begin{description}
\item[AND array] An $(n-1)\times (n-1)$ array of AND gates to compute $p_{i,j}=x_i\wedge y_j$  ($1\leq i,j\leq n-1$) for all $i$ and $j$.
\item[adder array] An  $(n-1)\times n$ array of half and full adders are used to compute $S$ from $p_{i,j}$'s.
\end{description}
Figure~\ref{fig:grade-school} shows the multiplication circuit for two 4-bit numbers.
It uses $3\times 3=9$ AND gates, 4 half adders, and 8 full adders.
It is evident that $(n-1)^2$ AND gates, $n$ half adders and $n^2-2n$ full adders are used for multiplication of two $n$-bit numbers.

\begin{figure}[!ht]
\centering
\includegraphics[scale=1]{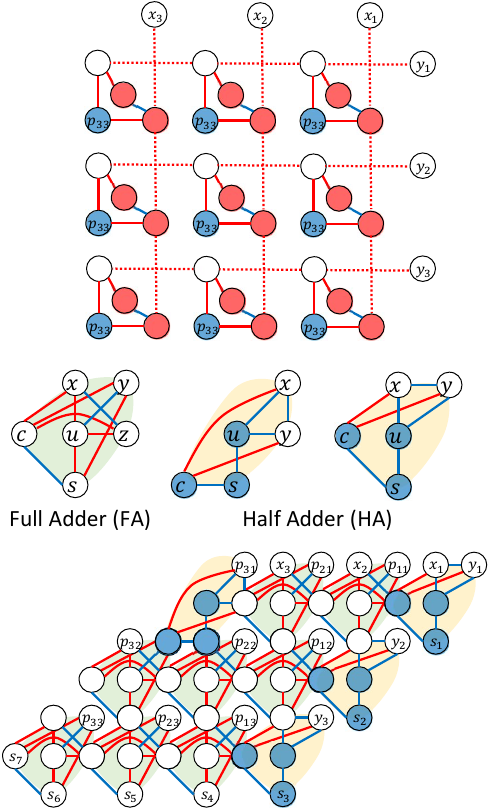}
\caption{Illustrating Ising models for an array of $3\times 3$ AND gates to compute $p_{i,j}$'s and
the $3\times 4$ array for half and full adders to compute the sum of $p_{i,j}$'s}
\label{fig:mul}
\end{figure}

We can derive Ising models for half and full adders through linear programming models in the same manner.
Table~\ref{tab:basic} includes the resulting zero-input unit Ising models $H^{zu}_{\rm HA}(x,y,s,c,u)$ and
$H^{zu}_{\rm FA}(x,y,z,s,c,u)$ for half and full adder, respectively.
Similarly to XOR gate, the same Ising models are obtained even if QUAD\_NUM is prioritized.
Also, interestingly, $H^{zu}_{\rm FA}(x,y,z,s,c,u)$ has no linear term.

Figure~\ref{fig:mul} depicts an implementation of the grade-school multiplication for two 4-bit numbers using an Ising model.
To implement a $3\times 3$ array of AND gates, we employ unit Ising models $H^{u}_{\rm AND}(a,b,x,u)$.
The connected components indicated by red dotted edges for each $y_j$ can be consolidated into one, because all nodes have no color.
However, those corresponding to red dotted edges for each $x_i$ cannot be merged due to red nodes.
If we utilize zero-input Ising models $H^{zu}(a,b,x,u)$ instead, they could be merged since all input nodes lack color.
Nevertheless, merging nodes would result in an Ising model with a dense topology, making it challenging to lay out in a two-dimensional plane.
Considering the layout perspective, it might be more favorable to employ $H^{u}_{\rm AND}(a,b,x,u)$ for implementing an AND array of size $(n-1)\times (n-1)$
and not to simplify it by merging nodes.
The figure also shows an Ising model for an adder array of size $3\times 4$ using $H^{zu}_{\rm HA}(x,y,s,c,u)$ and $H^{zu}_{\rm FA}(x,y,z,s,c,u)$.
It computes the sum of $p_{i,j}$'s, $x_i$'s and $y_j$'s appropriately to obtain the product $S$.
Since Ising models for half and full adders are zero-input unit, all red dotted edges connecting them can be merged.
The figure shows the resulting unit Ising model after merging  all red dotted edges connecting them.

In the forward computation of this multiplication Ising model, $X$ and $Y$ are set in the input nodes.
An optimal solution of the Ising model finds the product $S=X\cdot Y$ in the output nodes.
In the backward computation, the product of two prime numbers are set in the output nodes.
An optimal solution of the Ising model finds prime factors of $S$.

We will discuss the area complexity of the Ising model for multiplication and factorization over $n$ bits.
It is important to note that the Ising model comprises $O(n^2)$ nodes (or variables) and $O(n^2)$ edges (or quadratic terms)
From a practical standpoint, we explore the area complexity of the Ising model.
For this analysis, we adopt the assumptions used in the VLSI complexity model~\cite{Ullman84}.
Specifically, we assume that a qubit cell occupies $O(1)$ area and that the width of each interconnect is also $O(1)$.
The AND and adder arrays shown in Figure~\ref{fig:mul} are planar, and the length of each edge is $O(1)$.
Furthermore, the AND array and the adder array can be overlaid when arranged in a two-dimensional plane.
Consequently, the area complexity of the Ising model for multiplication and factorization over $n$ bits is $O(n^2)$, and the ASUQA for this can be accommodated in an area of size $O(n^2)$.
In digital circuit design, it has been proven~\cite{Brent81} that $AT^2 = \Omega(n^2)$ always holds for any $n$-bit multiplication circuit, where $A$ is the circuit area and $T$ is the multiplication time.
Following the same logic as in~\cite{Brent81}, we argue that an Ising model for $n$-bit multiplication and factorization requiring a single iteration of ideal annealing must occupy an area of size $\Omega(n^2)$ in a two-dimensional plane.
Intuitively, the graph bisection width of any Ising model for $n$-bit multiplication must be at least $\Omega(n)$, as $\Omega(n)$-bit information must be transferred from one side to the other of the bisection to complete $n$-bit multiplication.
If the bisection width is $o(n)$, it implies that there exists at least one pair of numbers for which the Ising model does not yield the correct multiplication or factorization result. 
Since a graph with bisection width $B$ necessitates an area of $\Omega(B^2)$ to be laid out in a two-dimensional plane, this indicates that no asymptotically smaller model of size $o(n^2)$ is possible,
thereby affirming the area optimality of our Ising model for $n$-bit multiplication and factorization.

\section{Related works}
\label{sec:related}

\subsection{QUBO model}
A Quadratic Unconstrained Binary Optimization (QUBO) model is a quadratic formula of binary values, each taking values of either $0$ or $1$.
Ising models and QUBO models can be converted to each other equivalently\cite{Tao20}.
Since handling $0/1$ binary values is easier and more intuitive than $-1/+1$ spin values, the D-Wave API, called \emph{dimod}~\cite{dimod-BQM},
for quantum annealing accepts QUBO models and automatically converts them into equivalent Ising models.
QUBO models have been investigated more often than Ising models, and several related works have targeted QUBO models.
For example, QUBO Solvers using GPUs~\cite{Okuyama19,Yasudo-JPDC22,Nakano23} and FPGAs~\cite{Matsubara17,Goto19,Kagawa21,Goto21} have been developed.
Thus, we will explain how a QUBO model can be converted to an equivalent Ising model.
Let $E(X)$ be a QUBO model with binary variables $X=x_0,x_1, \ldots, x_{n-1}$ defined by the following quadratic formula:
\begin{align*}
E(X) &= \QUAD{\sum_{i<j} W_{i,j} x_i x_j}~\LIN{+\sum_{i} W_{i} x_i},
\end{align*}
where $W_{i,j}$'s and $W_i$'s are integer coefficients of quadratic and linear terms.
The conversion to an equivalent Ising model follows.
Let $s_i$ be a spin variable satisfying $x_i={s_i+1\over 2}$.
Then, we can derive quadratic formula of spin variables equal to $E(X)$ as follows:
\begin{align*}
E(X) &= \QUAD{\sum_{i<j} W_{i,j} {(s_i+1)(s_j+1)\over 4}}~\LIN{+\sum_{i} W_{i}{s_i+1\over 2}}\\
       &= \QUAD{\sum_{i<j} {W_{i,j}\over 4}s_is_j}~\LIN{+\sum_{i}\left({W_i\over 2}+\sum_{i<j}{W_{i,j}\over 4}+\sum_{j<i}{W_{j,i}\over 4}\right)s_i}\\
   &\qquad +\sum_{i<j}{W_{i,j}\over 4}+\sum_{i}{W_i\over 2}
\end{align*}
By multiplying the right-hand side of the above formula by 4, we can obtain an equivalent Ising model with integer coefficients.
This conversion do not change the quadratic terms coefficients, but the coefficient of each linear term in the Ising model:
$
\LIN{\sum_{i}({W_i\over 2}+\sum_{i\neq j}{W_{j,i}\over 4})}
$,
can be quite large.
Thus, we should design Ising models to obtain those with the smallest maximum absolute value of coefficients, which can be accommodated by resolution-limited quantum annealers.

\subsection{Ising models for simulating logic gates}
For the reader's benefit, we first demonstrate how Ising models for certain logic gates can be obtained through manual calculation.
Specifically, we illustrate the derivation of an Ising model for simulating an AND gate.
Let $h_{\alpha,\beta,\gamma}(a,b,x)$ ($\alpha,\beta,\gamma\in\{-1,+1\}$) be a function defined as follows:
\begin{align*}
h_{\alpha,\beta,\gamma}(a,b,x) &= (1+\alpha a)(1+\beta b)(1+\gamma x) \\
& = \alpha\beta\gamma abx+\alpha\beta ab + \alpha\gamma ax+\beta\gamma bx+\alpha a+\beta b+\gamma x +1
\end{align*}
Clearly, $h_{\alpha,\beta,\gamma}(a,b,x)= 8$ if and only if all of $a=\alpha$, $b=\beta$, and $x=\gamma$ are satisfied; otherwise, $h_{\alpha,\beta,\gamma}(a,b,x)=0$.
To derive an Ising model for simulating the AND gate, we select a tuple of $\alpha,\beta,\gamma$ such that $\gamma\neq \alpha\wedge \beta$, as follows:
\begin{align*}
h_{-1,-1,+1}(a,b,x)
% & =(1-a)(1-b)(1+x) \\
&=  +abx\QUAD{+ab-ax-bx}~\LIN{-a-b+x}+1 \\
h_{-1,+1,+1}(a,b,x)
% & =(1-a)(1+b)(1+x)\\
& =  -abx\QUAD{-ab-ax+bx}~\LIN{-a+b+x}+1\\
h_{+1,-1,+1}(a,b,x) 
%& =(1+a)(1-b)(1+x)\\
& =  -abx\QUAD{-ab+ax-bx}~\LIN{+a-b+x}+1 \\
h_{+1,+1,-1}(a,b,x)
% & =(1+a)(1+b)(1-x) \\
& =  -abx\QUAD{+ab-ax-bx}~\LIN{+a+b-x}+1 
\end{align*}
The sum of these functions constitutes an Ising model model that takes the minimum value of 0 if and only if  $x=a\wedge b$.
However, due to the presence of a cubic term $abx$, it does not qualify as an Ising model.
To obtain an Ising model, we must eliminate this cubic term, which can be achieved by multiplying $h_{-1,-1,+1}$ by 3:
\begin{align*}
H(a,b,x)&=3h_{-1,-1,+1}+h_{-1,+1,+1}+h_{+1,-1,+1}+h_{+1,+1,-1}\\
&=\QUAD{ab-2ax-2bx}~\LIN{-a-b+2x}+3
\end{align*}
We can immediately derive $H_{\text{AND}}(a,b,x)$ from $H(a,b,x)$.
Similarly, an Ising model for simulating OR gate can be obtained in the same manner as follows:
\begin{align*}
H(a,b,x)&=h_{-1,-1,+1}+h_{-1,+1,-1}+h_{+1,-1,-1}+3h_{+1,+1,-1}\\
&=\QUAD{ab-2ax-2bx}~\LIN{+a+b-2x}+3
\end{align*}
In~\cite{Bian10}, exactly the same Ising models are shown although the paper did not mention how these formulas are obtained.
Moreover, Ising models for simulating all 16 logic functions have been shown~\cite{Su16} by extending the Ising models shown in~\cite{Bian10}.

Interestingly, it is not possible to apply the same method for deriving an Ising model for simulating XOR gate.
For XOR gate, we need to select a tuple of $\alpha,\beta,\gamma$ such that $\gamma\neq \alpha\oplus\beta$, as follows:
\begin{align*}
h_{-1,-1,-1}(a,b,x) &=-abx\QUAD{+ab+ax+bx}~\LIN{-a-b-x}+1\\
h_{-1,+1,+1}(a,b,x) &=-abx\QUAD{-ab-ax+bx}~\LIN{-a+b+x}+1\\
h_{+1,-1,+1}(a,b,x) &=-abx\QUAD{-ab+ax-bx}~\LIN{+a-b+x}+1\\
h_{+1,+1,-1}(a,b,x) &=-abx\QUAD{+ab-ax-bx}~\LIN{+a+b-x}+1
\end{align*}
Since all cubic terms $-abx$ have negative coefficients, it is not possible to remove it to obtain a quadratic formula by manual calculation.
Actually, as we have shown, XOR gate cannot be simulated by Ising model without ancillary variables.
The following Ising model $H(a,b,x,u)$ using an ancillary variable $u$ obtained by manual calculation has been presented in~\cite{Bybee23}:
\begin{align*}
H(a,b,x,u)
 &= 2u(-a-b+x+1) +a(b-x-1)+b(-x-1)+x
\end{align*}
This Ising model has 6 quadratic terms and 4 linear terms, and the maximum absolute value of coefficients is 2.
Thus, it shares the same characteristics as $H_{\rm XOR}(a,b,x,u)$ in Table~\ref{tab:basic}.
However, manually deriving $H^{ZU}_{\rm XOR}(a,b,x,u,v)$ is quite challenging.

In~\cite{Chang19}, a QUBO model for simulating an XOR gate using the ancillary variable $u$ is presented.
The QUBO model is defined as follows:
\begin{align*}
E(a,b,x,u) &= \QUAD{2ab-2ax-4au-2bx-4bu+4xu}~\LIN{+a+b+x+4u}
\end{align*}
They derive this QUBO model though a linear matrix inequality problem for finding all coefficients of QUBO models.
A linear inequality $E(a,b,x,u)>\mu$ is created if $x\neq a\oplus b$.
Further, when $x=a\oplus b$, they consider two options: (A) $E(a,b,x,0)=\mu$ and $E(a,b,x,1)>\mu$
or (B) $E(a,b,x,0)>\mu$ and $E(a,b,x,1)=\mu$.
They selected one of the two options by a trial-and-error method.
With 16 possible selections for four tuples $(a,b,x)$ satisfying $x=a\oplus b$, they chose one feasible solution for all the selections.
The resulting Ising model for XOR, obtained by converting the QUBO model $E(a,b,x,u)$ above, is equal to $H_{\rm XOR}(a,b,x,u)$ in Table~\ref{tab:basic}.
In contrast, our method, employing integer linear programming with objectives, is more sophisticated and capable of finding zero-input unit Ising models by objectives.

\subsection{QUBO models for multiplication and factorization}
Higher Order Unconstrained  Binary Optimization (HUBO) models, which contain higher order terms beyond quadratic terms,
 are sometimes utilized to derive a QUBO model~\cite{Jun23}.
For instance, a third order term $-abc$ can be transformed into linear and quadratic terms using an ancillary variable $u$ as demonstrated below:
\begin{align*}
E(a,b,c,u) &= -u(a+b+c-2)
\end{align*}
Here, $\min(E(a,b,c,0),E(a,b,c,1))$ attains a minimum value of $-1$ if $abc=1$, and $\min(E(a,b,c,0),E(a,b,c,1))=0$ if $abc \neq 1$.
Similarly, a third order term $+abc$ can be converted using the following expression:
\begin{align*}
E(a,b,c,u) &= u(a+b+c-1)+(ab+ac+bc)-(a+b+c)
\end{align*}
We can confirm that $\min(E(a,b,c,0),E(a,b,c,1))$ takes a minimum value $0$ if $abc=1$, and
$\min(E(a,b,c,0),E(a,b,c,1))=-1$ if $abc\neq 1$.
It can be verified that $\min(E(a,b,c,0),E(a,b,c,1))$ reaches a minimum value of $0$ when $abc=1$, and $\min(E(a,b,c,0),E(a,b,c,1))=-1$ if $abc \neq 1$.
Hence, third order terms can be reformulated into linear and quadratic terms through the introduction of an ancillary variable.
While fourth order terms such as $abcd$ can also undergo conversion, the process is notably more intricate.
For a detailed explanation of this conversion process, please refer to~\cite{Jun23}.

A HUBO model for the prime factorization of $Z=z_{2n-1}z_{2n-2}\cdots z_0$ into two $n$-bit primes $X=x_{n-1}x_{n-1}\cdots x_0$ and $Y=x_{n-1}x_{n-1}\cdots x_0$
can be directly derived from the following formula:
\begin{align*}
E(X,Y,Z)&=(\sum_{i=0}^{n-1}2^ix_i\cdot \sum_{i=0}^{n-1}2^iy_i - \sum_{i=0}^{2n-1}2^nz_i)^2
\end{align*}
It is evident that $E(X,Y,Z)$ reaches its minimum value of 0 when the product of $X$ and $Y$ equals $Z$.
The resulting HUBO model obtained by expanding this formula consists of fourth-order terms of $x_i$'s and $y_i$'s,
and thus can be converted into a QUBO model~\cite{Jun23}.
However, it is apparent that it contains coefficients with considerably large absolute values.

The column based algorithm for the grade-school method in Figure~\ref{fig:grade-school} to design QUBO model for factorization has been investigated~\cite{Jiang18,Wang20,Wang22}.
The idea is to design a HUBO model for each column of  the grade-school method with carry bits as follows:
The binary values of $x_1$ and $y_1$ can be derived by the following HUBO model:
\begin{align*}
E(x_1,y_1,c_{1,1})&=((2(x_1+y_1)+1)-(4c_{1,1}+2s_1+1))^2,
\end{align*}
where $c_{1,1}$ is an ancillary variable representing a carry to the next column.
Next, HUBO model for $x_2$ and $y_2$ can be described as follows:
\begin{align*}
\MoveEqLeft{E(x_1,x_2,y_1,y_2,c_{1,1},c_{2,1},c_{2,2})}\\
&= ((x_2+x_1y_1+y_2+c_{1,1})-(4c_{2,2}+2c_{2,1}+s_2))^2,
\end{align*}
where $4c_{2,1}+2c_{2,2}$ represents the carry value to the next column.
Following this method, HUBO models for upper bits can be derived similarly.
Summing all HUBO models thus obtained yields a HUBO model for factorization.
By minimizing the resulting values of the HUBO model, we can find an optimal solution.
This HUBO model involves fourth-order terms, which can be converted into quadratic terms to obtain a QUBO model for factorization.
The resulting QUBO model typically contains at least $O(n^3)$ quadratic terms, as each column necessitates $O(n^2)$ quadratic terms.
Moreover, arranging the QUBO model topology in a 2-dimensional plane necessitates $O(n^4)$ area complexity for $O(n^2)$ quadratic terms, leading to at least $O(n^5)$ area complexity.
In contrast, our Ising model for multiplication and factorization requires only $O(n^2)$ area complexity.
Additionally, Ising models based on this approach often feature large absolute values in coefficients.
For instance, an Ising model presented in~\cite{Wang22} for the factorization of $1630729 = 1277 \times 1277$ exhibits linear term coefficient ranges of $[-3554, 1983]$.

The idea of utilizing Ising models for factorization by implementing a binary multiplier based on the grade-school method might seem naive, yet a similar approach has been adopted in previous works~\cite{Andriyash16,Lanthaler23,Ding24}.
In~\cite{Andriyash16}, Ising models for AND, HA (half adder), and FA (full adder) gates are presented, with a multiplier implemented on the D-Wave 2000Q quantum annealer.
However, these Ising models contain more quadratic terms. For instance, the models for HA and FA have 8 and 13 quadratic terms, respectively, whereas our Ising models for them utilize only 7 and 10 quadratic terms.
Additionally, the Ising model for the AND gate requires quadratic terms with coefficients of $-2$ and $+2$.
The Ising models for factorization using a binary multiplier, as presented in~\cite{Andriyash16}, assume that the model can incorporate higher-order terms.
This suggests that factorization can be considered as satisfying a large set of Boolean functions. 
In paper~\cite{Ding24}, a full adder is defined as logic formula and then Ising model for factorization is defined through Ising models for satisfiability of Boolean functions presented in~\cite{Bian20}.
This approach, which leverages Boolean satisfiability, does not focus on the values of coefficients and thus requires coefficients with large absolute values.
They successfully embedded a $21\times 12$-bit multiplier in the D-Wave Advantage quantum annealer, which has 5760 qubit cells. 
However, since minor embedding is used, it necessitates large absolute values for biases and chain strengths.

\section{Conclusion}
\label{sec:concl}
In this study, we have introduced unit Ising models characterized by the property that every non-zero coefficient is either $-1$ or $+1$, along with zero-input unit Ising models specifically designed for simulating logic gates, wherein each linear term coefficient related to a gate input is set to zero.
We have demonstrated that these Ising models can be efficiently designed through integer linear programming.
This novel approach enables the optimization of Ising models for simulating not only logic gates but also circuit modules.
For a given combinational logic circuit that computes a Boolean function $f$, we can design an Ising model capable of computing both $f$ and its inverse $f^{-1}$ using this framework.
Owing to the fact that all coefficients of the resulting Ising model are constrained to $-1$ or $+1$, the model exhibits robust tolerance against flux noise during the quantum annealing process.
As a significant case study, we have demonstrated the construction of an Ising model capable of breaking RSA cryptography through factorization.
We further demonstrate that this model of area $O(n^2)$ is area-optimal, as any Ising model for multiplication and factorization must occupy at least $\Omega(n^2)$ area. 
Additionally, we introduced the concept of Application-Specific Unit Quantum Annealers (ASUQAs) for factorization, proposing a potentially powerful method for compromising current public-key cryptosystems with quantum computation.
\bibliographystyle{plain}
\bibliography{algorithm,gpu,fpga,sort,mining,halftone}

\begin{thebibliography}{10}

\bibitem{Gurobi}
Gurobi optimization.
\newblock https://www.gurobi.com.

\bibitem{Andriyash16}
Evgeny Andriyash, Zhengbing Bian, Fabian Chudak, Marshall Drew-Brook, Andrew~D.
  King, William~G. Macready, and Aidan Roy.
\newblock Boosting integer factoring performance via quantum annealing offsets.
\newblock Technical report, {D-Wave Systems}, 2016.

\bibitem{Bian20}
Zhengbing Bian, Fabian Chudak, William Macready, Aidan Roy, Roberto Sebastiani,
  and Stefano Varotti.
\newblock Solving sat (and maxsat) with a quantum annealer: Foundations,
  encodings, and preliminary results.
\newblock {\em Information and Computation}, 275:104609, 2020.

\bibitem{Bian10}
Zhengbing Bian, Fabian Chudak, William~G. Macready, and Geordie Rose.
\newblock The {Ising} model: teaching an old problem new tricks.
\newblock Technical report, {D-Wave Systems}, 2010.

\bibitem{DWaveAdvantage19}
Kelly Boothby, Paul Bunyk, Jack Raymond, and Aidan Roy.
\newblock Next-generation topology of {D-Wave} quantum processors.
\newblock techreport 14-1026A-C, D-Wave Systems, February 2019.

\bibitem{Brent81}
R.~P. Brent and H.~T. Kung.
\newblock The area-time complexity of binary multiplication.
\newblock {\em J. of ACM}, 28(3):521–534, jul 1981.

\bibitem{Bybee23}
Connor Bybee, Denis Kleyko, Dmitri~E. Nikonov, Amir Khosrowshahi, Bruno~A.
  Olshausen, and Friedrich~T. Sommer.
\newblock Efficient optimization with higher-order {Ising} machines.
\newblock {\em Nature Communications}, 14:6033, 2023.

\bibitem{Cai14}
Jun Cai, William~G. Macready, and Aidan Roy.
\newblock A practical heuristic for finding graph minors.
\newblock \url{https://arxiv.org/abs/1406.2741}, 2014.

\bibitem{Chang19}
Tyler~H. Chang, Thomas C.~H. Lux, and Sai~Sindhura Tipirneni.
\newblock Least-squares solutions to polynomial systems of equations with
  quantum annealing.
\newblock {\em Quantum Information Processing}, 18:374, 2019.

\bibitem{dimod-BQM}
{D-Wave Systems}.
\newblock dimod - {D-Wave} {Ocean} {Software} {Documentation}.
\newblock
  \url{https://docs.ocean.dwavesys.com/en/stable/docs\_dimod/reference/index.html}.

\bibitem{Ding24}
Jingwen Ding, Giuseppe Spallitta, and Roberto Sebastiani.
\newblock Effective prime factorization via quantum annealing by modular
  locally-structured embedding.
\newblock {\em Scientific Reports}, 14:3518, February 2024.

\bibitem{Goto21}
Hayato Goto, Kotaro Endo, Masaru Suzuki, Yoshisato Sakai, Taro Kanao, Yohei
  Hamakawa, Ryo Hidaka, Masaya Yamasaki, and Kosuke Tatsumura.
\newblock High-performance combinatorial optimization based on classical
  mechanics.
\newblock {\em Science Advances}, 7(6), February 2021.

\bibitem{Goto19}
Hayato Goto, Kosuke Tatsumura, and Alexander~R. Dixon.
\newblock Combinatorial optimization by simulating adiabatic bifurcations in
  nonlinear hamiltonian systems.
\newblock {\em Science Advances}, 5(4), April 2019.

\bibitem{Jiang18}
Shuxian Jiang, Keith~A. Britt, Alexander~J. McCaskey, Travis~S. Humble, and
  Sabre Kais.
\newblock Quantum annealing for prime factorization.
\newblock {\em Scientific Reports}, 8:17667, 2018.

\bibitem{Jun23}
Kyungtaek Jun and Hyunju Lee.
\newblock {HUBO} and {QUBO} models for prime factorization.
\newblock {\em Scientific Reports}, 13:10080, 2023.

\bibitem{Kadowaki98}
Tadashi Kadowaki and Hidetoshi Nishimori.
\newblock Quantum annealing in the transverse {Ising} model.
\newblock {\em PHYSICAL REVIEW E}, E58(5):5355--5363, November 1998.

\bibitem{Kagawa21}
Hiroshi Kagawa, Yasuaki Ito, Koji Nakano, Ryota Yasudo, Yuya Kawamata, Ryota
  Katsuki, Yusuke Tabata, Takashi Yazane, and Kenichiro Hamano.
\newblock High-throughput {FPGA} implementation for quadratic unconstrained
  binary optimization.
\newblock {\em Concurrency and Computation: Practice and Experience}, page
  e6565, August 2021.

\bibitem{Lanthaler23}
Martin Lanthaler, Benjamin~E. Niehoff, and Wolfgang Lechner.
\newblock Scalable set of reversible parity gates for integer factorization.
\newblock {\em Communications Physics}, 6:73, April 2023.

\bibitem{Levin03}
L.~A. Levin.
\newblock The tale of one-way functions.
\newblock {\em Problems of Information Transmission}, 39:92--103, 2003.

\bibitem{Lucas14}
Andrew Lucas.
\newblock Ising formulations of many {NP} problems.
\newblock {\em Frontiers in Physics}, 2(5), 2014.

\bibitem{Matsubara17}
Satoshi Matsubara, Hirotaka Tamura, Motomu Takatsu, Danny Yoo, Behraz
  Vatankhahghadim, Hironobu Yamasaki, Toshiyuki Miyazawa, Sanroku Tsukamoto,
  Yasuhiro Watanabe, Kazuya Takemoto, and Ali Sheikholeslami.
\newblock Ising-model optimizer with parallel-trial bit-sieve engine.
\newblock In {\em Proc. of International Conference on Complex, Intelligent,
  and Software Intensive Systems}, pages 432--438, 2017.

\bibitem{Advantage}
Catherine McGeoch and Pau Farr\'{e}.
\newblock The {D-Wave} {Advantage} {System}: An overview.
\newblock Technical report, D-Wave Systems, 2020.

\bibitem{McGeoch19}
Catherine~C. McGeoch, Richard Harris, Steven~P. Reinhardt, and Paul Bunyk.
\newblock Practical annealing-based quantum computing.
\newblock {\em IEEE Computer}, 52:38--46, June 2019.

\bibitem{Mead89}
Carver Mead.
\newblock {\em Analog {VLSI} and Neural Systems}.
\newblock Addison-Wesley, 1989.

\bibitem{Murty87}
Katta~G. Murty and Santosh~N. Kabadi.
\newblock Some {NP}-complete problems in quadratic and nonlinear programming.
\newblock {\em Mathematical Programming}, 39:117--129, 1987.

\bibitem{Nakano23}
Koji Nakano, Daisuke Takafuji, Yasuaki Ito, Takashi Yazane, Junko Yano, Shiro
  Ozaki, Ryota Katsuki, and Rie Mori.
\newblock Diverse adaptive bulksearch: a framework for solving {QUBO} problems
  on multiple {GPUs}.
\newblock In {\em Proc. of International Parallel and Distributed
  ProcessingSymposium Workshops}, pages 314--325. Proc. of International
  Parallel and Distributed ProcessingSymposium Workshops, May 2023.

\bibitem{Nakano23-dual}
Koji Nakano, Shunsuke Tsukiyama, Yasuaki Ito, Takashi Yazane, Junko Yano,
  Takumi Kato, Shiro Ozaki, Rie Mori, and Ryota Katsuki.
\newblock Dual-matrix domain wall: A novel technique for generating
  permutations by {QUBO} and {Ising} models with quadratic sizes.
\newblock {\em Technologies}, 11(5), Oct. 2023.

\bibitem{Okuyama19}
Takuya Okuyama, Tomohiro Sonobe, Kenichi Kawarabayashi, and Masanao Yamaoka.
\newblock Binary optimization by momentum annealing.
\newblock {\em PHYSICAL REVIEW E}, 100(1):012111, July 2019.

\bibitem{RSA}
R.~L. Rivest, A.~Shamir, and L.~Adleman.
\newblock A method for obtaining digital signatures and public-key
  cryptosystems.
\newblock {\em Communications of the ACM}, 21:120 -- 126, 1978.

\bibitem{Su16}
Juexiao Su, Tianheng Tu, and Lei He.
\newblock A quantum annealing approach for boolean satisfiability problem.
\newblock In {\em Proceedings of the 53rd Annual Design Automation Conference},
  2016.

\bibitem{Tao20}
Masaki Tao, Koji Nakano, Yasuaki Ito, Ryota Yasudo, Masaru Tatekawa, Ryota
  Katsuki, Takashi Yazane, and Yoko Inaba.
\newblock A work-time optimal parallel exhaustive search algorithm for the
  {QUBO} and the {Ising} model, with {GPU} implementation.
\newblock In {\em International Parallel and Distributed Processing Symposium
  Workshops}, pages 557--566. Proc. of International Parallel and Distributed
  Processing Symposium Workshops, May 2020.

\bibitem{Ullman84}
Jeffrey~D. Ullman.
\newblock {\em Computational Aspects of VLSI}.
\newblock Computer Science Press, 1984.

\bibitem{Wang20}
Baonan Wang, Feng Hu, Haonan Yao, and Chao Wang.
\newblock Prime factorization algorithm based on parameter optimization of
  ising model.
\newblock {\em Scientific Reports}, 10:7106, 2020.

\bibitem{Wang22}
Baonan Wang, Xiaoting Yang, and Dan Zhang.
\newblock Research on quantum annealing integer factorization based on
  different columns.
\newblock {\em Frontiers in Physics}, 10, 2022.

\bibitem{Yasudo-JPDC22}
Ryota Yasudo, Koji Nakano, Yasuaki Ito, Ryota Katsuki, Yusuke Tabata, Takashi
  Yazane, and Kenichiro Hamano.
\newblock {GPU}-accelerated scalable solver with bit permutated cyclic-min
  algorithm for quadratic unconstrained binary optimization.
\newblock {\em Journal of Parallel and Distributed Computing}, 167:109--122,
  September 2022.

\bibitem{Zaborniak21}
Tristan Zaborniak and Rog\'{e}rio de~Sousa.
\newblock Benchmarking {Hamiltonian} noise in the {D-Wave} quantum annealer.
\newblock {\em IEEE Transactions on Quantum Engineering}, January 2021.

\end{thebibliography}

\end{document}